# Prediction of Human Performance Capability during Software Development using Classification

Sangita Gupta[1], Suma V.[2]

[1]Jain University, Bangalore Dept of CSE, Bangalore, India
sgjain.res@gmail.com
[2] RIIC, Dayanada Sagar Institute, Bangalore, DBIT, Bangalore, India
sumavdsce@gmail.com

**Abstract.** — The quality of human capital is crucial for software companies to maintain competitive advantages in knowledge economy era. Software companies recognize superior talent as a business advantage. They increasingly recognize the critical linkage between effective talent and business success. However, software companies suffering from high turnover rates often find it hard to recruit the right talents. There is an urgent need to develop a personnel selection mechanism to find the talents who are the most suitable for their software projects. Data mining techniques assures exploring the information from the historical projects depending on which the project manager can make decisions for producing high quality software. This study aims to fill the gap by developing a data mining framework based on decision tree and association rules to refocus on criteria for personnel selection. An empirical study was conducted in a software company to support their hiring decision for project members. The results demonstrated that there is a need to refocus on selection criteria for quality objectives. Better selection criteria was identified by patterns obtained from data mining models by integrating knowledge from software project database and authors research techniques.

## 1. Introduction

Human aspect of software engineering has become one of the main concerns in software companies to achieve quality objectives. Software industries are now paying attention to select the right talent who can perform consistently throughout all generic framework activities and execute the process properly. Software quality depends on people and process quality during development[10]. Hence, this paper proposes the use of data mining algorithms that can exploit the patterns in the historical data and predict the performance based on project personnel attributes and thereby enhance the process and quality of software.

Data mining is a new and promising field for knowledge discovery. Data mining is the process of extracting knowledge from data [8]. It uses a combination of an explicit knowledge base, sophisticated analytical skills, domain knowledge to uncover hidden trends and patterns. These trends and patterns can be extracted on by using various data mining algorithms. To create a model, the algorithm first analyzes a large set of data and finds specific patterns. The algorithm then uses the results of the analysis to define the parameters of the mining model. These parameters are then applied across the entire data set. Subsequently, patterns and detailed statistics can be extracted. Through classification, one can identify association rules. Categorization uses rule induction algorithms to handle categorical outcomes, such as good, average and poor as in this study. There are a wide range of available algorithms for such purpose. Many of them are implemented in WEKA [7]. WEKA is a cluster of machine learning algorithms for data mining tasks. The algorithms can be applied directly to a dataset. WEKA contains large variety of tools for data pre-processing, classification, regression, clustering, association rules, and visualization.

The paper is organized as follows: Section 2 provides the related work and background of data mining Algorithms. Section 3 presents the research methodology to derive at a conclusion while Section 4 depicts the obtained result. Section 5 discusses the summary of this paper and its future scope.

## 2. Related work and background

The growing complexities of software and increasing demand of software projects have led to the progress of continual research in the areas of effective project management. Data mining has proven as one of the established techniques for effective project management recently. Data mining methodologies are developed for several applications including various aspects of software development. It works on large quantities of data to discover meaningful patterns and rules. Authors in [4] surveyed different data mining algorithms used for defect prediction in software and also discuss the performance and effectiveness of data mining algorithms. Authors of [5] made a comparative analysis of performance of Density-Based Spatial Clustering of Applications with Noise (DBSCAN) and Neuro-Fuzzy System for prediction of level of severity of faults present in Java based object oriented software system. Data which comprises of project personnel data provides a rich resource for knowledge discovery and decision support[9]. Data mining results in decision through methods and not assumptions. In [13] authors have done an empirical study for selection criteria for software industry by introducing a knowledge based decision tree algorithm. Authors in [2] have worked on the improvement of employee selection, by building a model, using data mining techniques. Depending on few selected attributes, the model could predict their performance. Some of these attributes are personal characteristics, educational and professional attributes. They specified age, gender, marital status, experience, education, major subjects and school tires as potential factors that might affect the performance. As a result for their study, they found that employee performance is highly affected by education degree, the school tire, and the job experience. The authors in [3] searched on certain factors that affect the job performance. They reviewed previous studies, experience, salary, training, working conditions and job satisfaction on the performance. As a result of their research, it was found that several factors affected the employee's performance such as position of the employee in the company, working conditions and environment. Highly educated and qualified employees showed dissatisfaction of bad working conditions and thus affected their performance negatively. Employees of low qualifications, on the other hand, showed high performance in spite of the bad conditions. Experience showed positive relationship in most cases, while education did not yield clear relationship with the performance. Data mining thus supports various techniques including statistics, decision tree, genetic algorithm, bayes classification and visualization techniques for analyses and prediction. It further deals with association, clustering and classification [1].

This part of the research therefore involves applying data in WEKA tool and derives a classification model for selection criteria. Some of the data mining algorithms are ID3, C4.5 and CART [12]. Decision trees are a hierarchical structure with leaves and stems. The hierarchical structure of decision trees represents different levels of attributes. Every leaf reveals the classification of an attribute, while the stems indicate the conditions of the attributes. Given training set, a decision tree can be constructed depending on various methods to provide valuable information about the attributes and their patterns [6]. The authors of [6] have developed ID3 (Iterative Dichotomise 3) which is based on Hunts algorithm. The tree is constructed in two phases namely tree building and pruning. Authors have thus developed C4.5 which is a successor to ID3 and is based on Hunt's algorithm [11]. Classification And Regression Trees (CART) is yet another popularly available algorithm for WEKA users. CART was introduced by Breiman. It handles both categorical and continuous attributes to build a decision tree in addition to handle missing values. ID3 and C4.5 algorithms have multiple branches, however CART produces binary splits and thereby binary tree. WEKA contains tools for regression, classification, clustering, association rules and visualization. The classify panel enables the user to apply classification algorithms to the resulting dataset, estimate the accuracy of the resulting predictive model, visualize erroneous predictions and the model itself as shown in section IV of this study. The further section will discuss more about research methodology and results obtained from ID3, CART and C4.5.

## 3. Research Methodology

The main objective of the study is to build a performance model of an employee based on his/her attributes.

This investigation focused upon the selection criteria of right project personnel that yield effective results for better software quality through a prediction technique. The objective was to concentrate on human aspect of software engineering by selecting the appropriate people.

This research therefore aimed to construct a framework for human resource data mining to explore the relationships between personnel profiles and its effect on software development. Through the proposed methodology, hidden information could be extracted from large volumes of personnel data and thus the project leaders are able to comprehend and focus on selection for the software project through the discovered knowledge. Fig. 1 shows the framework

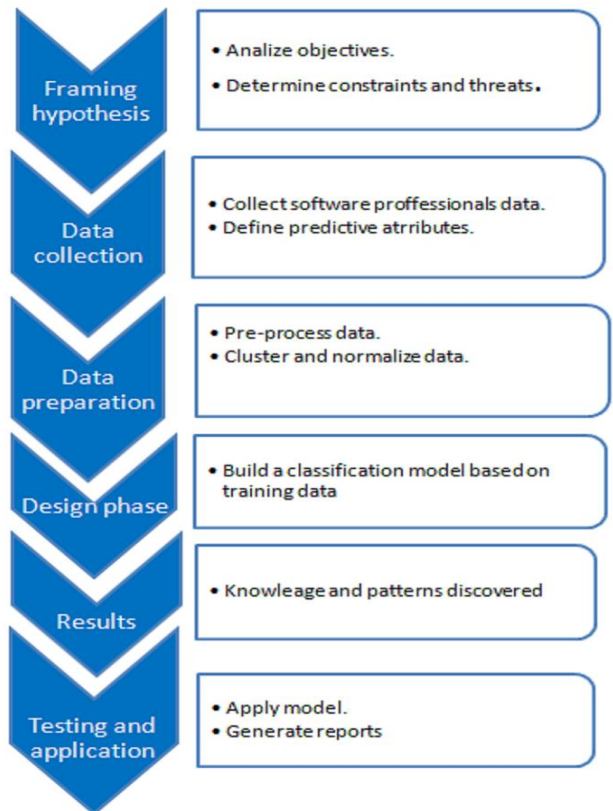

**Fig 1- Research Framework**

Figure 1 indicates the methodology followed for this research. Having made a deep study of literature survey and industry based investigations, this research focused upon the formulation of hypothesis.

   a.  **Hypothesis** –Project personnel with certain attributes values will perform similarly.

Project personnel information is collected from those projects which are developed in similar domain using common technology and programming language. The constraint of this study is that it has dealt with short term projects of duration two to three years with windows operating system and C/C++ as programming language for web based applications.

   b.  **Data Collection**

Data was collected in collaboration with the project team. Modes of data collection included brainstorming, discussions, interviews, and argumentation. The project leaders provided goals and experts in human resources suggested criteria. The most relevant questions considered were
- How to select the relevant attributes
- How to structure them

In order to collect the required data, a questionnaire was prepared and given to both project leaders and project personnel. Information regarding various attributes was asked in the questionnaire that might predict the performance class. Attributes involving personal, educational and work related details were collected. However, this study aimed at educational, internal assessment results and work related attributes.

   c.  **Data Preparation**

Having obtained the responses for the questionnaires, the process of preparing the data was accomplished. The list of collected and selected attributes which is relevant for this research is specified below.

PS- Programming skill. This variable was obtained through internal assessment by the company. It was mapped into three categories as good, average and poor based on the marks out of 100. Good for 75% and above, average for 60 to 75% and poor below 60%. The three categories with the above mentioned grading were taken to map the input with the output or target.

RS- Reasoning skill. During selection company takes various assessments. This variable was one of the internal assessments, which is categorized similar to PS.

DKA-Domain Knowledge Assessment. This is yet another level of internal assessment made by the company. It was categorized and normalized similar to PS and RS.

TE-Time Efficiency. It was obtained from the project data through project leaders. It was obtained in the form of YES and NO.

GPA- General Percentile Assessment. This is obtained from the database pertaining to personal attributes of candidate. It is mapped into good (for >7.5), average (for <7.5 an >6.5) and poor for (<6.5)

CS-Communication Skills. The value for this variable is fetched from the project leaders. This is mapped into good, average and poor.

P- Performance. This is the target or output class. The value for performance variable is acquired from project team leaders in terms of good, average or poor, which is based on the quality of software developed. The company has a evaluation system every month and the consolidated results for the tenure of the project was considered. Sample instances of the training data is shown in Table 1.

**TABLE 1- Portion of Training Data Set**

| S.No. | GPA | DKA | PS | TE | CS | RS | P |
|---|---|---|---|---|---|---|---|
| 1 | Good | Good | Good | Yes | Good | Good | Good |
| 2 | Good | Good | Average | No | Good | Good | Good |
| 3 | Good | Good | Average | No | Average | Average | Average |
| 4 | Good | Average | Good | Yes | Good | Good | Average |
| 5 | Good | Average | Average | Yes | Good | Good | Average |
| 6 | Good | Poor | Poor | No | Average | Poor | Poor |

d. **Implementation of Mining Model**

Based on the background and related work, a training set with the attributes depicted in data preparation is selected to test against their effectiveness on the employee performance.

The algorithm used for classification in this study is ID3, C4.5 and CART. Under the "Test options", the 10-fold cross-validation is selected for the evaluation approach. Since, there is no separate evaluation data set, this option was necessary to get a reasonable idea of accuracy of the generated model. The model is generated in the form of decision tree as shown in the results section. These predictive models provide analytical way to formulate selection criteria.

## 4. Results

The three decision trees of predictive models obtained from the training data set by three machine learning algorithms: the ID3 decision tree algorithm, the CART decision tree algorithm and the C4.5 also called J48 in WEKA environment are shown in TABLE 2, TABLE 3 and TABLE 4.

Table 2 infers the decision tree obtained from ID3 in the tree option of classification in WEKA tools. Table 3 depicts the binary decision tree obtained in run information for CART option in tree option of classification. It shows the result with root node again being PS. Since it is a binary tree it has two branches with PS as poor and not poor. Table 4 shows the C4.5 pruned tree in J48 option of tree option in classification. Table 5 shows the accuracy of ID3, C4.5 and CART algorithms for classification applied on the above data sets using 10-fold cross validation.

**TABLE 2 - ID3 Decision Tree**

| |
|---|
| === Classifier model (full training set) === |
| Id3 |
| PS = Good<br>\| RS = Good: Good<br>\| RS = good: Good<br>\| RS = Average: Good<br>\| RS = Poor: Average<br>\| RS = poor: null<br>PS = Average<br>\| DKA = Good: Average<br>\| DKA = Average: Average<br>\| DKA = Poor<br>\| \| GPA = Good: Good<br>\| \| GPA = Average: Average<br>\| \| GPA = Poor: Poor<br>PS = Poor: Poor |

Table 2 , result obtained from run information of WEKA tool, clearly indicates that if PS is good and RS is good or average, the performance is good. Also generally when PS is good then RS is either good or average and performance is good. If PS is average then performance is average. However, if PS is average but if GPA is good then there exists a possibility of performance being good. When PS is poor then his performance is poor irrespective of all other attribute values.

**TABLE 3- Cart Decision Tree**

| | |
|---|---|
| === Classifier model (full training set) === CART Decision Tree    Size of the Tree: 5 | Number of Leaf Nodes: 3 |
| PS=(Poor): Poor(13.0/0.0)<br>PS!=(Poor)<br>\| PS=(Average)\|(Poor): Average(12.0/2.0)<br>\| PS!=(Average)\|(Poor): Good(12.0/1.0) | |

Table 3 infers that if PS is poor then the training set showed all 13 records with performance as poor. There were 13 records in the training set with PS poor and all showed poor performance. In the second level CART tree once again split PS  into average and good. When PS value is average then 12 records showed average performance and two records showed poor performance. When PS is good then 12 records showed good performance and only one showed not good. Table 3 also indicates that PS is most dominating attribute. The best match is when all poor show poor and all average show average and all good show good performance. However, compared to other attributes, PS showed the best match and thereby to deem the attribute with highest importance.

**TABLE 4 - C4.5/J48 Pruned Tree**

| | |
|---|---|
| === Classifier model (full training set) ===J48 pruned tree--    tree: 4 | Number of Leaves: 3Size of the |
| PS = Good: Good (13.0/1.0)<br>PS = Average: Average (14.0/2.0)<br>PS = Poor: Poor (13.0) | |

Table 4, run information of J48 depicts that if PS is good then out of 13 records all mapped to performance as good except 1. If PS is average then apart from 2 records all indicated an average performance. However, when PS is poor all records showed poor performance. These decision trees also provide interesting insights into hidden patterns in the project personnel performance by showing the importance of attributes in a hierarchical manner.  The tree indicated that  programming skills, domain knowledge and reasoning skills are more relevant than college aggregate. The tree generated using the C4.5 algorithm also indicated that the programming skills is most effective attribute for an IT company.

TABLE 5- Comparison Of Techniques

| Algorithm | Correctly Classified Instances | Incorrectly Classified Instances |
|---|---|---|
| ID3 | 90.5% | 7.5% |
| CART | 92.5% | 7.5 % |
| C4.5 | 90.5% | 7.5% |

After building the tree, WEKA tool again cross validates the result obtained with the data set and gives the information about correctly and incorrectly classified records. Table 5 indicates that ID3 algorithm could classify 90.5% of data correctly. CART and C4.5 showed similar results by classifying 92.5% and 90.5% correctly. CART showed highest correctly classified instances.

Since all three options showed similar results we stopped our investigation for other classification model in WEKA tools.

Thus, this investigation indicates that performance of project member is good if programming skill and reasoning skills are good irrespective of his college aggregate or communication skills.

## 5. Conclusion

In this paper data mining is used to predict the performance of software project member on the basis of previous database or training set. This paper has focused on the human aspect of software engineering to achieve good quality of software by building a classification model for predicting employees' performance based on certain attributes. Data mining techniques could identify those attributes required in a project member which will contribute to good performance and thereby enhance software quality and success. Classification techniques like ID3, CART and C4.5 showed similar results. Performance was earlier assumed to be best for candidates having a good college aggregate.

However, this study showed that other talent attributes like programming skills and reasoning skills have proved to be more important, although software companies emphasize upon aggregate percentile and followed the same trend for many years. Due to lack of analytical method in human aspects, software companies were not selecting the right people who could perform well in the software process and thereby failed to achieve the desired quality in the time and cost constraints.

Data mining techniques have given very interesting patterns and helped in earlier identification of project members who will perform well. This study enables the managers to refocus on human capability criteria and thereby enhance the development process of software project. Just like all process within generic framework of software development is given importance, human aspect also needs a deeper investigation for effective software development. Without the right people even the best process analysis and development are bound to fail.

Further scope of this study is to investigate on several projects of different domains and take into account more attributes of project personnel and correlate it with software quality and success.